\documentclass[9pt,twocolumn,twoside]{pnas-new}
\templatetype{pnasresearcharticle} % Choose template 
\usepackage{bm}
\usepackage{color}
\usepackage{amsmath, amsthm, amssymb}

\newcommand{\exref}[1]{\ref{#1}}

\renewcommand{\eqref}[1]{equation~\ref{#1}}
\newcommand{\eqsref}[1]{equations~\ref{#1}}

\newcommand{\eqsand}[2]{equations \ref{#1} and~\ref{#2}}

\newcommand{\figref}[1]{figure~\ref{#1}}

\newcommand{\bea}{\begin{eqnarray}}
\newcommand{\eea}{\end{eqnarray}}
\newcommand{\beq}{\begin{equation}}
\newcommand{\eeq}{\end{equation}}
\newcommand{\lt}{\left}
\newcommand{\rt}{\right}

\newcommand{\la}{\langle}
\newcommand{\ra}{\rangle}

\newcommand{\mbf}[1]{{\bf #1}}

\newcommand{\rmd}{\mathrm{d}}
\newcommand{\dd}{\partial}
\newcommand{\vdel}{\boldsymbol{\nabla}}
\newcommand{\vdperp}{\vdel_\perp}
\newcommand{\dperp}{\nabla_\perp}
\newcommand{\dpar}{\nabla_\parallel}

\newcommand{\vr}{\mbf{r}}
\newcommand{\vrperp}{\vr_\perp}
\newcommand{\vz}{\hat{\mbf{z}}}

\newcommand{\vk}{\mbf{k}}

\newcommand{\vkperp}{\vk_\perp}
\newcommand{\kperp}{k_\perp}

\newcommand{\kpar}{k_\parallel}

\newcommand{\lperp}{\lambda}
\newcommand{\lpar}{\ell_\parallel}
\newcommand{\mfp}{\lambda_\mathrm{mfp}}

\newcommand{\vv}{\mbf{v}}

\newcommand{\vperp}{v_\perp}
\newcommand{\vpar}{v_\parallel}

\newcommand{\vth}{v_{{\rm th}}}

\newcommand{\vA}{v_{\rm A}}

\newcommand{\vZ}{\mbf{Z}}
\newcommand{\vu}{\mbf{u}}

\newcommand{\vE}{\mbf{E}}
\newcommand{\vB}{\mbf{B}}
\newcommand{\vb}{\mbf{b}}

\newcommand{\dn}{\delta n}
\newcommand{\dB}{\delta\! B}

\renewcommand{\tt}{\tilde t}

\title{Fluidization of collisionless plasma turbulence}

\author[a,b,1]{Romain Meyrand}
\author[c,d]{Anjor Kanekar}
\author[c,e]{William Dorland}
\author[e,f]{Alexander A.\ Schekochihin} 
\affil[a]{LPP, \'Ecole Polytechnique, F-91128 Palaiseau Cedex, France}
\affil[b]{Space Sciences Laboratory, University of California, Berkeley, CA 94720, USA}
\affil[c]{Department of Physics, University of Maryland, College Park, Maryland 20742-3511, USA}
\affil[d]{Palantir Technologies, 20 Soho Square, London W1D 3QW, UK}
\affil[e]{Rudolf Peierls Centre for Theoretical Physics, University of Oxford, Clarendon Laboratory, Parks Road, Oxford OX1 3PU, UK}
\affil[f]{Merton College, Oxford OX1 4JD, UK}

\leadauthor{Meyrand} 

\significancestatement{Two textbook physical processes compete to thermalize 
turbulent fluctuations in collisionless plasmas: Kolmogorov's ``cascade'' 
to small spatial scales, where dissipation occurs, and Landau's damping, which 
transfers energy to small scales in velocity space via ``phase mixing'', also 
leading to dissipation. We show that, in a magnetized 
plasma, another textbook process, plasma echo, brings energy back 
from phase space and on average cancels the effect of phase mixing. Energy 
cascades effectively as it would in a fluid system and thus Kolmogorov 
wins the competition with Landau for the free energy in a collisionless turbulent 
plasma. This reaffirms the universality of Kolmogorov's picture of turbulence 
and explains, for example, the broad Kolmogorov-like spectra of density 
fluctuations observed in the solar wind.}

\authorcontributions{R.M., A.A.S., W.D., and A.K.\ conceived this project, 
including its theoretical rationale, and interpreted the results; 
A.K.\ carried out some preliminary numerical experiments and their analysis; 
R.M.\ wrote a new code and carried out the numerical simulations 
and data analysis presented in the paper; 
R.M., A.A.S., and W.D.\ wrote the paper.}
\authordeclaration{The authors declare no conflict of interest.}
\correspondingauthor{\textsuperscript{1}To whom correspondence should be addressed. E-mail: romain.meyrand@lpp.polytechnique.fr}


\begin{abstract}
In a collisionless, magnetized plasma, particles may stream freely
along magnetic-field lines, leading to ``phase mixing'' of their distribution function 
and consequently to smoothing out of any ``compressive'' fluctuations 
(of density, pressure, etc.,). 
This rapid mixing underlies Landau damping of these fluctuations 
in a quiescent plasma---one of the most 
fundamental physical phenomena that make plasma different from a conventional fluid. 
Nevertheless, broad power-law spectra of compressive fluctuations are observed 
in {\it turbulent} astrophysical plasmas (most vividly, in the solar wind) 
under conditions conducive to strong Landau damping. 
Elsewhere in nature, such spectra are normally associated with fluid turbulence, 
where energy cannot be dissipated in the inertial scale range and is therefore cascaded 
from large scales to small. By direct numerical simulations 
and theoretical arguments, it is shown here that turbulence of compressive 
fluctuations in collisionless plasmas strongly resembles one 
in a collisional fluid and does have broad power-law spectra. 
This ``fluidization'' of collisionless plasmas 
occurs because phase mixing is strongly suppressed on average by ``stochastic echoes'', 
arising due to nonlinear advection of the particle distribution by turbulent motions.  
Besides resolving the long-standing puzzle of
observed compressive fluctuations in the solar wind, our results 
suggest a conceptual shift for understanding kinetic plasma turbulence generally: 
rather than being a system where Landau damping plays the role of dissipation, 
a collisionless plasma is effectively dissipationless except at very small scales. 
The universality of ``fluid'' turbulence physics is thus reaffirmed even for a kinetic, 
collisionless system.
\end{abstract}

\dates{This manuscript was compiled on \today}
\doi{\url{www.pnas.org/cgi/doi/10.1073/pnas.XXXXXXXXXX}}

\begin{document}

% Optional adjustment to line up main text (after abstract) of first page with line numbers, when using both lineno and twocolumn options.
% You should only change this length when you've finalized the article contents.
\verticaladjustment{-2pt}

\maketitle
\thispagestyle{firststyle}
\ifthenelse{\boolean{shortarticle}}{\ifthenelse{\boolean{singlecolumn}}{\abscontentformatted}{\abscontent}}{}

\dropcap{W}hat makes plasma turbulence a particularly fascinating
subject is, apart from its ubiquity in the laboratory and in space,
its kinetic nature. Plasma dynamics are six-dimensional, 
evolving in the phase space of positions and velocities, $(\vr, \vv)$. Free energy 
that is injected at large scales must be transferred to 
small scales in $\vr$ or $\vv$ before it can be
dissipated (see \cite{sch08} and references therein). 
In nature, the range of scales implicated
can be quite large, and the general theory is, therefore, a formidable challenge. 
At spatial scales larger than the Larmor radii of the particles, on time scales
longer that those particles' Larmor periods, and with account taken of 
the smallness of the electron mass relative to the ion 
mass, turbulent, magnetized plasma can be described by a reduced version \cite{sch09} of the
drift-kinetic, or ``Kinetic-Magnetohydrodynamic'' 
\cite{kulsrud83}, approximation, whereby the phase space 
is reduced to four dimensions $(\vr, \vpar)$, where only the velocity $\vpar$ parallel 
to the magnetic field survives as a kinetic variable. Even this simplified
problem is manifestly kinetic in its nature, conceptually interesting and as yet
unsolved. 

In this regime, the transfer of free energy to small scales in phase space
can be conceptualized as a superposition of two fundamental processes:
{\em nonlinear} spatial mixing of the fluctuating electric and magnetic fields, as well as  
of the particle distribution function, by the turbulent $\vE\times\vB$ flows 
across the magnetic field, and {\em linear} phase mixing of the
distribution function, brought about by particle streaming along magnetic-field
lines. The former is {\em turbulence} in
the usual ``fluid'' sense \cite{K41}. The latter, in the linear
plasma theory, is known as {\em Landau \cite{landau46} damping:}
it involves the transfer of free energy from the the ``fluid'' quantities
such as density, velocity, magnetic field, etc., to
higher velocity moments of the perturbed distribution function
(fine-scale structure in velocity space; see, e.g.,
\cite{hammett90,hammett92,kanekar15}).  How the ``turbulent
cascade'' works in the presence of these two types of mixing is a
fundamentally interesting question, which, with the advent of
high-resolution measurements of turbulence in the solar wind
\cite{chen16,servidio17} 
and of kinetic simulations of this turbulence
\cite{howes08prl,howes11prl,told15,cerri18,groselj18prl,franci18,kawazura18ie}, 
has increasingly preoccupied theoreticians and modellers (indeed, not
only in application to space plasmas, but also, for an even longer
time, to fusion ones
\cite{hammett92,weiland92,hammett93,dorland93,beer96,hatch14,sch16,parker16}). 

Because the computational resources required to solve the full, 
four-dimensional system (as we shall do here) 
are formidable and difficult to extend 
to realistic situations, several groups have taken a middle road, 
replacing the distribution function with a few of its $\vpar$ moments.  
One must close the resulting fluid hierarchy in some fashion. 
``Landau-fluid'' closures 
\cite{hammett90,hammett92,dorland93,beer96,snyder97,snyder01,passot04,sulem15,tassi16} 
enforce outgoing boundary conditions for 
free energy in the linearized equations, from resolved moments to unresolved. 
They thus model the transfer of energy to small scales in 
velocity space assuming that the energy carried to the higher resolved moments is ultimately 
dissipated by (very weak) collisions. 
An even more radically pragmatic modelling choice, popular in astrophysical 
applications, is to assume that the
effect of phase mixing is merely to damp the spatial part of the turbulent
cascade of the ``fluid'' quantities at a scale-dependent rate equal to
the linear Landau damping rate in the system \cite{howes08jgr,podesta10kaw,howes10,passot15,kunz18}. 
We shall see that the latter approach in general misses an essential property of turbulent 
plasma---but we shall give some credence to the notion that Landau-fluid models, carefully 
constructed, might capture the relevant physics. 

In the context of solar-wind turbulence (and, more generally, of
collisionless plasma turbulence, of which the solar wind is a
particularly well-diagnosed instance), the question of how 
the turbulent fluctuations' energy (free energy) is thermalized must be tackled
if one is to explain the measured spectra of
``compressive'' (density and magnetic-field-strength)
perturbations. In the fluid (short-mean-free-path) magnetohydrodynamic (MHD) theory, 
these perturbations correspond to the slow-wave and entropy modes. 
Compressions are supported in MHD by collisions,
which prevent particles from freely streaming through and away, 
along the magnetic field. Because momentum and energy are 
conserved in each collision, the inertial-range fluctuations are 
undamped (viscosity and resistivity do dissipate 
them, but at much smaller spatial scales). Instead, 
they are passively advected by the Alfv\'enic
perturbations, including in situations when the latter are turbulent \cite{lithwick01,sch09}. 
As passive tracers, compressive MHD perturbations should 
have a spectrum that follows the spectrum of 
the Alfv\'enic turbulence---and indeed solar-wind measurements show that they do
\cite{celnikier83,celnikier87,marsch90,bershadskii04,hnat05,kellogg05,chen11}.
However, the solar-wind plasma at 1 AU, where these measurements are
done, is essentially collisionless: its mean free path is approximately 1 AU.  
In such a plasma, compressive
perturbations, while still passive with respect to the Alfv\'enic
ones \cite{sch09}, are in fact subject to Landau damping (known in 
this context as Barnes \cite{barnes66} damping) 
at rates characterized by free streaming along field lines.  
Thus, the variances of density and field-strength fluctuations 
are not conserved---they form part of the total compressive free energy, which 
includes also the variance of the perturbed ion distribution function and could 
be rapidly redistributed through all scales in
velocity space---equivalently, to higher velocity-space moments---until it is ultimately 
dissipated by collisions. By this 
conventional argument, the wavenumber spectra of the compressive perturbations 
should decay more steeply in a collisionless plasma than in a collisional one, 
because at each scale, energy is removed into phase
space at a rate at least similar to the rate at which it is passed to
the next smaller scale.\footnote{The damping rate of perturbations with parallel wave number 
$\kpar$ is $\sim |\kpar|\vth$, where $\vth$ is the ion thermal speed. The nonlinear cascade rate due to mixing of the compressive perturbations by the Alfv\'enic turbulence is, by the critical-balance conjecture \cite{GS95,GS97}, $\sim\kpar\vA$, where the Alfv\'en speed is typically $\vA\sim\vth$. 
The steepening of the spectra in a turbulent system where the damping rate is comparable to the 
cascade rate at every scale is explained in Sec.~2.4.4 of \cite{sch16}.} 
Yet, this is not observed. Not only do the observed compressive fluctuations 
have spectra that follow the Alfv\'enic fluctuations' spectra, 
as if the flow of free energy to higher moments were substantially blocked, 
they also display surprisingly ``fluid'' dynamics \cite{verscharen17}. 

A range of possible explanations for this ``fluid''
behaviour of a collisionless plasma have been mooted: e.g., that
Landau damping might be quantitatively weak \cite{lithwick01}, or
that compressive fluctuations, unlike the Alfv\'enic ones, do not 
develop small scales along the (perturbed) magnetic field lines, 
rendering the damping ineffective \cite{sch09}.
What in fact happens is subtler:
while the compressive fluctuations do have a parallel cascade and thus do
phase-mix vigorously, much of their energy flux into phase space due
to this phase mixing is on average canceled by a return flux from phase
space, due to the stochastic version \cite{sch16} 
of the plasma echo phenomenon \cite{gould67,malmberg68}. 
The result is effective suppression of Landau damping of the compressive fluctuations. 
With access to this loss channel inhibited, the 
density and magnetic-field-strength perturbations
instead develop increasingly sharp, small-scale spatial features, 
characterized by broad power-law wavenumber spectra. 
Their free energy is ultimately thermalized by processes that occur at 
small spatial scales (at and below the ion Larmor radius \cite{sch09}), 
outside the scope of this study.

\section*{Theoretical Framework}
Motivated by observational evidence that turbulence 
in the solar wind consists of low-frequency fluctuations (compared to the ion
Larmor frequency), with negligible energy in the fast magnetosonic
modes, we tackle the problem in the framework of ``Kinetic Reduced
Magnetohydrodynamics'' (KRMHD) \cite{sch09}, which is 
the long-wavelength limit of gyrokinetics \cite{frieman82,howes06,kunz18} 
and the anisotropic reduction of Kinetic MHD \cite{kulsrud83,kunz15}.  
The electrons are isothermal in this approximation, so only the ions require a kinetic treatment. 
The collisional limit of KRMHD is the well-known ``Reduced MHD'' \cite{kadomtsev74,strauss76}. 

The applicability of KRMHD is limited by a few constraints. Most
importantly, these include magnetized ions ($\lperp \gg \rho_i$,
where $\rho_i$ is the thermal ion Larmor radius and $\lperp$ is any 
target fluctuation's scale length measured across the magnetic field); 
spatial anisotropy ($\lperp \ll \lpar$, where $\lpar$ is a
fluctuation's scale length along the local magnetic field); 
and low frequency ($\omega \ll \Omega$, where $\omega$
represents any dynamical frequency of interest and $\Omega$ is the ion
Larmor frequency). The fluctuating fields must be small in
comparison to their background values (ordered $\sim\omega/\Omega\sim\lambda/\lpar$), 
and the background magnetic field is assumed to be (locally) straight, with constant magnitude. 
The ion mean free path $\mfp$ can be long or short compared to
$\lpar$ without violating KRMHD orderings. 
Similarly, the ratio of the sound speed to the Alfv\'en speed 
($\sim\sqrt{\beta_i}$, where $\beta_i$ is the raio of the ion thermal 
to magnetic energies) can be large or small. 
KRMHD is well suited to the study of the inertial-range turbulence, in 
the solar wind and elsewhere.

Under this approximation, Alfv\'enic fluctuations are completely
unaffected by the compressive fluctuations, and the compressive 
fluctuations are affected {\it only} by the Alfv\'enic fluctuations  
and not by one another \cite{sch09}. 

\subsection*{Alfv\'en waves}  
In KRMHD, Alfv\'enic fluctuations are purely transverse,
incompressible, nonlinearly interacting Alfv\'en-wave packets, 
which propagate in both directions along a straight, static 
background magnetic field $\vB_0 = B_0 \vz$, 
in a plasma with mean mass density $\rho_0$. 
Two scalar stream (flux) functions $\Phi = \Phi(\vr, t)$ and
$\Psi = \Psi(\vr, t)$ describe the fluctuating field-perpendicular flow velocity
$\vu = \vz \times \vdel \Phi$ and magnetic field
$\vb = \vz \times \vdel \Psi$, respectively; 
the latter is in units of the Alfv\'en speed
$\vA=B_0/\sqrt{4\pi\rho_0}$. 
Only the $z$ components of the vorticity $\vdel\times\vu$ and current $\vdel\times\vb$ are 
non-zero, equal to $\dperp^2 \Phi$ and $\dperp^2 \Psi$, respectively. 

The Els\"asser representation $\vZ^{\pm}=\vu\pm\vb$ \cite{elsasser50} 
brings to the fore key features of Alfv\'enic fluctuations in full MHD. 
In our reduced theory, 
the Els\"asser potentials $\zeta^\pm = \Phi \pm \Psi$ and ``vorticities'' 
$\omega^\pm = \dperp^2\zeta^\pm$ are useful. In terms of these functions, 
the KRMHD equations for Alfv\'enic fluctuations are
\beq
\lt(\frac{\dd}{\dd t} \mp v_A \frac{\dd}{\dd z}\rt) \omega^\pm = 
- \lt\{ \zeta^\mp, \omega^\pm \rt\}
- \lt\{ \dd_j \zeta^\mp, \dd_j \zeta^\pm \rt\},
\label{eq:KRMHDa}
\eeq
where 
$\lt\{f,g\rt\} \equiv \dd_x f\, \dd_y g - \dd_y f\, \dd_x g$. In the final term 
in \eqref{eq:KRMHDa}, summation over $j = x,y$ is implied. 

Important Alfv\'enic phenomena are easily read from \eqref{eq:KRMHDa}. 
The linear terms (on the left-hand side) describe 
perturbations propagating up and down the background 
magnetic field with speed $v_A$. According to the right-hand side, 
only counter-propagating perturbations interact. The 
compressive fluctuations do not appear.
Note that the local {\it direction} of the magnetic field is determined by the Alfv\'en 
waves, because $\vB/B \approx \vz + \vb/\vA$. KRMHD 
separately tracks fluctuations of the magnetic-field {\it strength}, 
$\dB$, as described below. 

\subsection*{Ion kinetics}  
Compressive fluctuations, namely those of density ($\dn$) and pressure, 
are calculated via moments of the ion distribution function 
perturbed from a Maxwellian equilibrium; the perturbed magnetic-field 
strength $\dB$ is obtained from these via pressure balance (across the 
mean field $\vB_0$). In KRMHD, as a result of 
the straight-$\vB_0$ equilibrium geometry, magnetic-moment conservation, 
and the restriction to long wavelengths ($\lperp\gg\rho_i$), 
one can integrate the perturbed distribution function over perpendicular velocities 
$\vperp$, retaining only the $\vperp^0$ and $\vperp^2$ moments, which are 
required to obtain $\dn$ and $\dB$. 
The ion kinetics are thus encoded by two kinetic scalar fields, 
$g = g^{(i)} (\vr, \vpar, t)$, $i=1, 2$, which are  
particular linear combinations of the $\vperp^0$ and $\vperp^2$ moments, 
chosen to produce two decoupled kinetic equations \cite{sch09}:
\beq
\frac{\rmd g^{(i)}}{\rmd t} + \vpar \dpar g^{(i)} + \vpar F_0  \dpar \phi^{(i)} = 0,
\label{eq:gb}
\eeq
where $F_0 = \exp{(-\vpar^2/\vth^2)}/\sqrt{\pi}\vth$ is the background 
Maxwellian distribution. Ions are accelerated along the moving field lines by 
the parallel electric field and the mirror force. For each $g^{(i)}$, 
the appropriate linear combination of these forces is 
accounted for by its potential $\phi^{(i)} = \alpha^{(i)}\int \rmd\vpar \, g^{(i)}(\vpar)$, 
where the constant prefactors $\alpha^{(i)}$ depend on plasma parameters---the ratios of ion thermal 
to magnetic energies (``plasma beta'', $\beta_i\equiv 8\pi n_i T_i/B_0^2$), 
of the ion to electron temperatures, $\tau\equiv T_i/T_e$, 
and of the ion to electron charge, $Z\equiv q_i/|e|$. 
The explicit expressions are 
$\alpha^{(i)} = (\tau/Z - 1/\beta_i \pm \kappa)^{-1}$,
with ``$+$'' for $i=1$ and ``$-$'' for $i=2$ and 
$\kappa = [(1+\tau/Z)^2 + 1/\beta_i^2]^{1/2}$. 
At any time and location, $\dn$ and $\dB$ can be determined as 
linear combinations of $\phi^{(1)}$ and $\phi^{(2)}$, with coefficients 
that also depend on the plasma parameters:
\begin{align}
\label{eq:dn}
\frac{\dn}{n_i} &= \frac{1}{2\kappa}\lt[\sigma\frac{\phi^{(1)}}{\alpha^{(1)}}
- \frac{2\tau}{Z\beta_i}\frac{\phi^{(2)}}{\alpha^{(2)}}\rt],\\
\frac{\dB}{B_0} &= \frac{1}{2\kappa}\lt[\sigma\frac{\phi^{(2)}}{\alpha^{(2)}} 
- \lt(1+\frac{Z}{\tau}\rt)\frac{\phi^{(1)}}{\alpha^{(1)}}\rt],
\label{eq:dB}
\end{align} 
where $\sigma = 1 + \tau/Z + 1/\beta_i + \kappa$. 
The particular forms of these coefficients or of $\alpha^{(i)}$
are not important for the forthcoming discussion, which 
hinges just on the mathematical structure of \eqref{eq:gb}. 
We shall drop the superscripts of $g$ and $\phi$ wherever this 
causes no ambiguity.  

The apparent simplicity of \eqref{eq:gb} is an intentional feature 
of the KRMHD model, but a few subtleties deserve mention. 
The equations for $g^{(1)}$ and $g^{(2)}$ are totally decoupled from
one another, but not from the background Alfv\'enic perturbations,
which move the plasma and the field lines around. This effect 
enters via the ``convective'' derivatives 
$\rmd/\rmd t\equiv\dd/\dd t +\vu\cdot\vdperp$ and
$\dpar \equiv \dd/\dd z + (\vb/\vA)\cdot\vdperp$. 

In the presence of Alfv\'enic fluctuations, since the magnetic field
lines are ``frozen'' into the plasma, the Alfv\'enic flows $\vu$ 
move the field lines $\vb$ and the plasma together.
Thus, \eqref{eq:gb} might seem to be a linear equation
that has been re-expressed in a well-defined Lagrangian frame \cite{sch09}. 
However, after about one eddy-turnover time, 
structure in $\vb$ develops at arbitrarily small spatial scales \cite{eyink15,eyink18,boozer18}. 
Thus, in the presence of any
resistivity (not shown in \eqref{eq:KRMHDa}, but included in our simulations and in reality), 
the field lines' identities and thus the Lagrangian frame are lost. 
Consequently, a density fluctuation (for example) that is 
momentarily aligned with the local magnetic field 
develops finer parallel wavelengths in about one eddy-turnover time.\footnote{Note that, absent resistivity, the parallel electric field associated with Alfv\'enic fluctuations is zero, and one might therefore mistakenly believe that the potentials $\phi$ describe all parallel electric fields also in resistive KRMHD. In fact, the moving field lines associated with the Alfv\'enic fluctuations can change their topology in the presence of resistivity and generate parallel electric fields in the process. In KRMHD, the compressive fluctuations cannot see these resistive parallel electric fields; they are not included in $\phi$. Thus, when field lines resistively reconnect, KRMHD compressive fluctuations are undisturbed, but get relabeled as the field-line identities change. It is this relabeling that causes the compressive fluctuations to inherit the parallel structure of the Alfv\'enic fluctuations.}
This process of field-line dissolution and replacement
manifests as a forward parallel cascade of compressive fluctuations by background 
Alfv\'enic fluctuations, as we observe in our simulations 
(see \figref{fig:par_cascade}). 

The derivative operators in \eqref{eq:gb} are independent of $\vpar$ and 
the $\vpar$ integrals extend from $-\infty$ to $+\infty$. 
These properties and the appearance of the Maxwellian function $F_0(\vpar)$ make 
Hermite polynomials a natural orthogonal basis for $g(\vpar)$. 
That is, one can decompose 
\beq
\label{eq:Hermite}
g(\vpar) = \sum_{m=0}^\infty \frac{H_m(\vpar/\vth) F_0(\vpar)}{\sqrt{2^m m!}}\,g_m,
%\quad
%g_m = \int_{-\infty}^{+\infty}\rmd\vpar\frac{H_m(\vpar/\vth)}{\sqrt{2^m m!}}\,g(\vpar), 
\eeq
where $H_m(x)=(-1)^m e^{x^2}(\rmd/\rmd x)^me^{-x^2}$ is the Hermite polynomial of degree $m$. 
It is convenient to describe the velocity-space structure of $g(\vpar)$ in terms of the spectral 
coefficients $g_m$. Just as a fluid cascade is described in terms of a flux of
energy from low to high wavenumbers, one may describe
collisionless damping as a flux of free energy from low $m$ to high $m$. 

In this context, it is appropriate to state explicitly what we mean by free energy \cite{sch09}. 
It follows immediately from \eqref{eq:gb} that it has a quadratic invariant: 
\begin{align}
\nonumber 
W &= \int\frac{\rmd^3\vr}{V}\lt(\int\rmd\vpar\frac{g^2}{2F_0} + \frac{\phi^2}{2\alpha}\rt)\\
&= \frac{1}{2}\sum_{\vk}\lt(\sum_{m=0}^\infty|g_{m,\vk}|^2 + \alpha|g_{0,\vk}|^2\rt), 
\label{eq:W}
\end{align}
where $V$ is the volume of the domain and 
the second expression follows from the Parseval theorem for the Fourier and 
Hermite functions and provides an explicit motivation for discussing free-energy 
flows in the $(m,\vk)$ phase space. It is not hard to show that $W$ is minus 
the entropy of the perturbed distribution plus the energy of the fluctuating 
magnetic field (up to a factor of $T_i$), which is what motivates the term 
``free energy''. 

\subsection*{Landau damping, phase mixing and stochastic echo} 
In the absence of
Alfv\'enic fluctuations, \eqref{eq:gb} is the one-dimensional,
unmagnetized problem originally considered by Landau \cite{landau46}, 
so the compressive fields $\phi$ are subject to his damping. 
In the presence of Alfv\'enic turbulence, the rate of collisionless damping
is {\it enhanced}, because the damping rate and
the associated rate of formation of structure in velocity space are
proportional to $|\kpar|\vth$, which is 
{\it increased} by the parallel cascade of compressive fluctuations. 
If this were the end of the story, i.e., if the compressive-fluctuation 
spectra were determined solely by simple ``superposition'' 
of the turbulent cascade and Landau damping, then 
one would predict steep (and possibly non-universal) Fourier spectra \cite{bratanov13prl,sch16} 
and shallow Hermite spectra \cite{zocco11,kanekar15}, corresponding 
to free energy spreading to small scales in velocity space as quickly as 
to small scales in real space. But there is more to the story.

In the absence of collisions, phase mixing is formally
reversible. Spatial perturbations of fields $\phi$ decay in time as spatial
perturbations of the distribution function are transferred to
finer scales in velocity space (higher Hermite moments $g_m$), 
but until collisions scramble the phases, 
the process can in principle be reversed.  Consider \eqref{eq:gb} for an initial value problem, 
with the initial perturbation $\phi(\vrperp,\ell) \sim \cos(\kpar \ell)$ along some
field line labeled by the perpendicular coordinate~$\vrperp$.
If we could Fourier transform $g(\vrperp, \ell, \vpar, t)$
along the moving field line, we would find the ballistic (or ``free-streaming'') 
part of the response to be $g_{\kpar} \sim \exp(-i\kpar \vpar t)$; 
the argument of the exponent is
the ``phase'' in ``phase mixing''. The ``mixing'' occurs when one
integrates $g_{\kpar}(\vpar)$ over $\vpar$, 
as one would to find a self-consistent $\phi$---the Maxwellian
velocity dependence of $F_0(\vpar)$ in the initial condition is
then mixed with the perturbation's advancing phase \cite{hammett90,hammett92}. 
A more complete analysis is required to recover 
Landau's full story, but the essential connection between phase mixing and the damping of fields
that is exposed here is authentic. Velocity-space integrals decay even
as $g_{\kpar}(\vpar)$ itself only becomes more
oscillatory in $\vpar$, without decaying. If time were reversed, the
accumulated phase would unmix, and the original $\phi$ perturbations
would grow in time, until the unmixing were complete. 

It is not easy to reverse time, but the scattering of a
compressive wave by an Alfv\'en wave that is propagating 
in the opposite direction along the field line
has the same mathematical effect: 
such an interaction causes the compressive wave's 
$\kpar$ to change sign, the phase subsequently 
unmixes and the field perturbation regenerates. The most familiar example 
of this phase unmixing is the textbook phenomenon of plasma echo \cite{gould67,malmberg68}. 
The idea of a stream of stochastic echoes produced by a 
sequence of nonlinear interactions has been the object of several
recent studies of {\it electrostatic} ($\vb=0$) plasma turbulence 
\cite{sch16,kanekar15phd,parker16,adkins18}.
Here, we explore for the first time the possibility of 
stochastic echoes in a plasma with moving field lines ($\vb\neq0$). 
As described above, the compressive fluctuations couple to 
the Alfv\'enic fluctuations in a nontrivial way, as the field lines along
which the particles stream are being advected by the same flows as 
the particles' distribution function $g$. 
Although there is as yet no complete theory for this problem, 
our numerical results suggest that elements of the electrostatic theory \cite{sch16,adkins18} 
are germane. This generalization opens the way to understanding kinetic 
turbulence in heliospheric (solar-wind)
and similarly collisionless astrophysical plasmas, which are usually 
well into the electromagnetic regime.

\begin{figure}[t]
\includegraphics[width=1.0\linewidth]{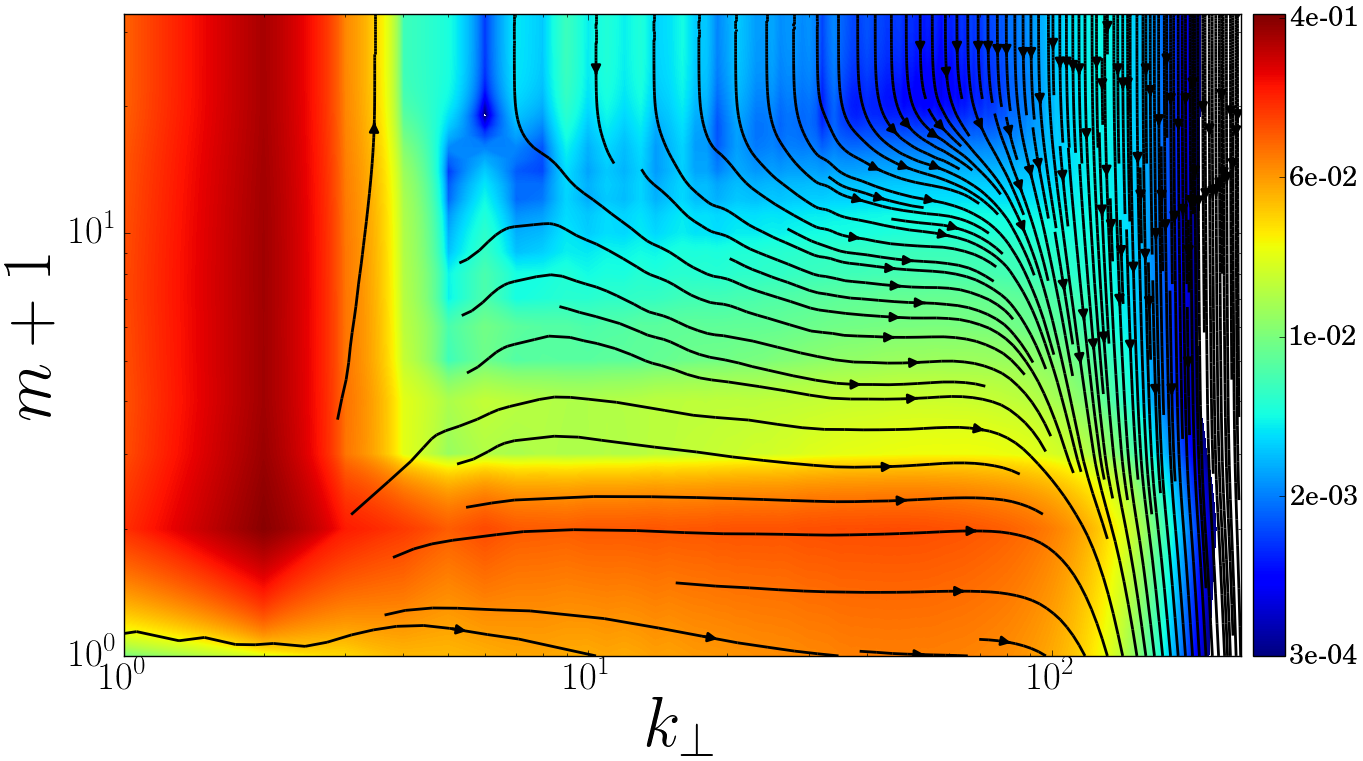}
\caption{The flux $(\Pi,\Gamma)$ (see \eqsand{eq:Gamma}{eq:Pi})
of free energy $W$ (\eqref{eq:W}) of the kinetic field $g^{(1)}$, 
normalized to its injection rate, as a function of the Hermite number $m$ (vertical) and 
perpendicular wavenumber $\kperp$ (horizontal). Colors represent the magnitude of the flux 
and arrows its streamlines. 
Away from the stirring (at small $m$ and small $\kperp)$ and damping 
(at large $m$ and/or large $\kperp$), the time-averaged net turbulent free-energy flux 
to high $m$ is very small; Landau damping is largely suppressed by stochastic echoes
(return flux from high to low $m$)
and so different Hermite moments $g_m$ of the ion distribution function are effectively 
energetically insulated from each other.} 
\label{fig:flux}
\end{figure}

Below we show evidence that nonlinear, stochastic echoes 
are present, and that they can have profound effects on 
observable quantities. Our key finding, 
best illustrated by \figref{fig:flux}, is that in the inertial range, 
the average net flux of free energy to smaller velocity-space 
scales (higher Hermite moments) is very small in comparison  
with the average net flux of free energy to smaller spatial 
scales (higher perpendicular wavenumbers). In the figure, 
this conclusion is supported by the nearly horizontal 
stream lines of the flux in the inertial range 
(away from the stirring at long wavelengths 
and from the dissipation at the smallest scales). After a brief discussion of 
the set up of our numerical experiment, we provide further evidence from
the simulations, including detailed spectra and structure 
functions such as can be (or have been) obtained from solar-wind measurements.

\section*{Numerical Experiment}
\subsection*{Numerical set up}
Our simulations track the evolution of driven and damped compressive 
fluctuations in a sea of driven and damped Alfv\'enic turbulence for a number 
of the latter's eddy-turnover times---well into the statistically stationary state, 
so statistical averages can be reliably calculated. 

We solve \eqsref{eq:KRMHDa} and \exref{eq:gb} spectrally in Fourier--Hermite space. 
Namely, we solve \eqref{eq:gb} in the form of a number of coupled 
``fluid'' equations for the Hermite moments $g_m$, defined by \eqref{eq:Hermite}:
\begin{align}
%\label{eq:g0}
%&\frac{\rmd g_0}{\rmd t} + \vth\dpar\frac{g_1}{\sqrt{2}} = 0,\\
%\label{eq:g1}
%&\frac{\rmd g_1}{\rmd t} + \vth\dpar\lt(g_2 + \frac{1+\alpha}{\sqrt{2}}\,g_0\rt) = 0,\\
\frac{\rmd g_m}{\rmd t} + \vth\dpar\!\lt[\sqrt{\frac{m+1}{2}}\,g_{m+1} 
+ (1+\alpha\,\delta_{m,1})\sqrt{\frac{m}{2}}\,g_{m-1}\rt] = 0 %\quad m\ge2,
\label{eq:gm} 
\end{align}  
and $\phi = \alpha g_0$. 
We advance these equations in time with a third-order modified Williamson algorithm (a
four-step, low-storage Runge-Kutta method). 
The spatial dependence of all fields is expressed as discrete Fourier series 
and we use the standard pseudo-spectral approach to solve the linear and nonlinear 
terms containing spatial derivatives. 
The simulation domain is a triply periodic cube of size $2\pi$ in each 
direction.\footnote{Note that in KRMHD, there is a rescaling symmetry whereby all 
fluctuation amplitudes 
relative to the corresponding background values (e.g., $\dB/B_0$, $u/\vA$, $g/F_0$ etc.) 
can be arbitrarily rescaled as long as the ratios of all perpendicular to parallel scales 
are rescaled by the same amount (i.e., the fluctuation amplitudes and $\kpar/\kperp$ 
are arbitrarily small). Therefore, the parallel and perpendicular units of length 
are independent.} The code units are set by this and by $\vA=1$. 
The nonlinear terms are partially dealiased using a phase-shift method \cite{patterson71}. 
The total number of
spectral modes is therefore smaller than the total number of mesh points used 
to evaluate the nonlinear terms; we report the spatial 
size of a given simulation in terms of the latter. 
The simulations presented here have 
spatial resolution from $256^3$ (\figref{fig:g_spec}) to $512^3$ (all figures). 
 
The description of compressive fluctuations provided by
\eqref{eq:gb} is appropriate for the inertial range, but the
{\it production} of fluctuations at large scales and their removal at small 
scales in a simulation require one to add forcing and dissipation, respectively.  
The forcing occurs at $m=1$ and low wavenumbers, viz., 
$|k_x|,|k_y|,|k_z| = 1, 2$.
The strength of the forcing can be arbitrary because 
\eqref{eq:gb} is linear in $g$ and so the 
amplitude of the compressive fluctuations is arbitrary. 
Our forcing is designed so as to keep the rate of injection of the compressive 
free energy exactly constant. 

We add two forms of dissipation to \eqref{eq:gb}, 
to absorb energy at small scales in
$\vr$ (``hyperviscosity'') and $\vpar$ (``hypercollisionality'') 
without changing the dynamics at large scales. 
We tested multiple specific forms of dissipation in the course of this
study. Our results require the presence of the dissipation at small scales, 
but do not depend on its specific form. In the simulations reported here, 
we add $-\mu\kperp^8g_m - \nu m^6g_m$ to the right-hand side of \eqref{eq:gm} with $m\ge2$. 
The values of $\mu$ and $\nu$ are chosen so that this dissipation is significant 
only for modes with the largest values of $\kperp$ and $m$ but removes energy 
without creating bottlenecks or reflections. 

In practice, the Hermite series of $g(\vpar)$ (\eqref{eq:Hermite}) 
is truncated at some $m=M$. In the absence of collisions, the Hermite 
\eqsref{eq:gm} fail to close, because the evolution of each $g_m$ 
depends on $g_{m+1}$ (and $g_{m-1})$. 
However, with the hypercollisional cutoff described above, one can 
take $g_{M+1} = 0$ to a very good approximation as long as
$M$ is chosen large enough. This is our approach. The simulations reported here 
have resolution from $M=32$ (all figures) to $M=128$ (\figref{fig:g_spec}). 

The Alfv\'en waves are also stirred and damped. We stir them 
by forcing at low wavenumbers (the same as for the kinetic field $g$) and in the velocity field only. 
The forcing is designed to keep the injected power $\epsilon$ exactly constant, 
while maintaining the rate of cross-helocity injection exactly zero \cite{teaca11}.
The magnitude of $\epsilon$ allows us to control the character of the resulting Alfv\'enic turbulence: 
weaker injection produces weakly interacting Alfv\'en waves, but increasing it 
pushes the system towards strong, critically balanced turbulence \cite{GS95,GS97}. 
This is the relevant regime that we study here. It is achieved when $\epsilon=1$ in code units. 
Like the kinetic field, the Alfv\'enic fields are also dissipated by eighth-order 
``hyperviscosity'' and ``hyperresistivity'', artificially set to be numerically equal.  

The plasma parameters $\beta_i$, $\tau$ and $Z$ are all set to unity. They 
enter via the constants $\alpha^{(i)}$ in the definitions of the potentials $\phi^{(i)}$ 
in terms of $g^{(i)}$ and affect their Landau-damping rates. 
They are also needed to compute the relative amplitudes of $\dn$ and $\dB$, 
via \eqsand{eq:dn}{eq:dB}.
The results presented here do not depend significantly upon these
as long as they are all $\sim1$. 

\begin{figure}[t]
\begin{center}
\includegraphics[width=0.85\linewidth]{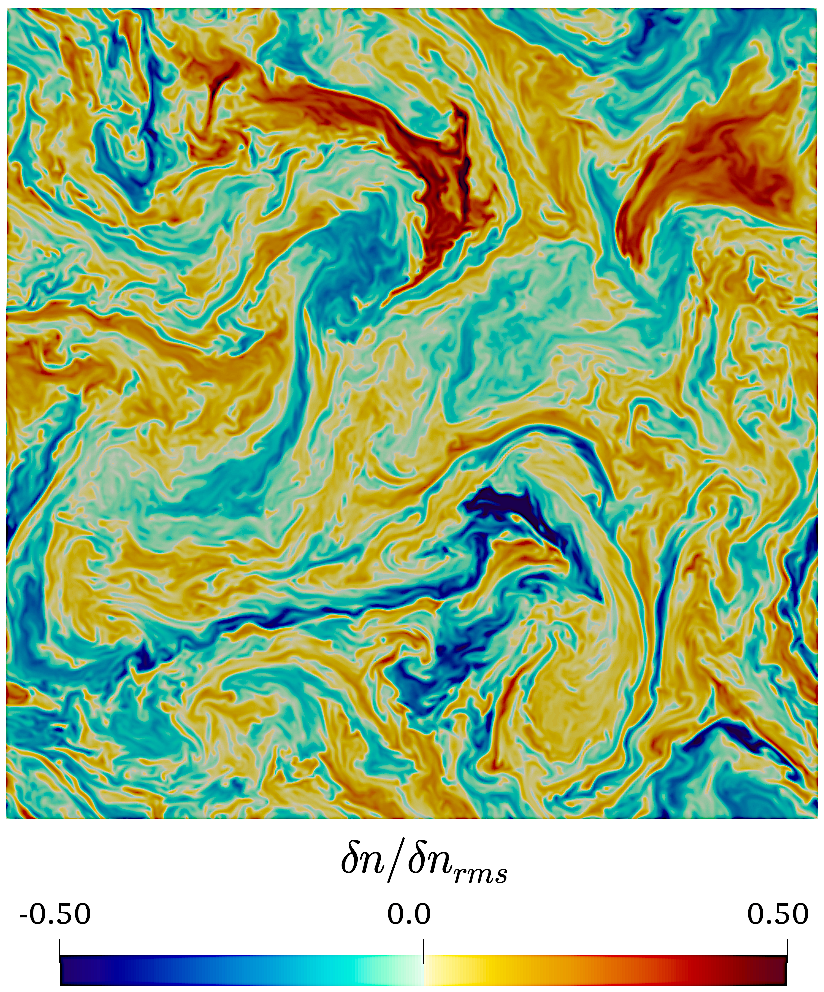}\\
\includegraphics[width=0.9\linewidth]{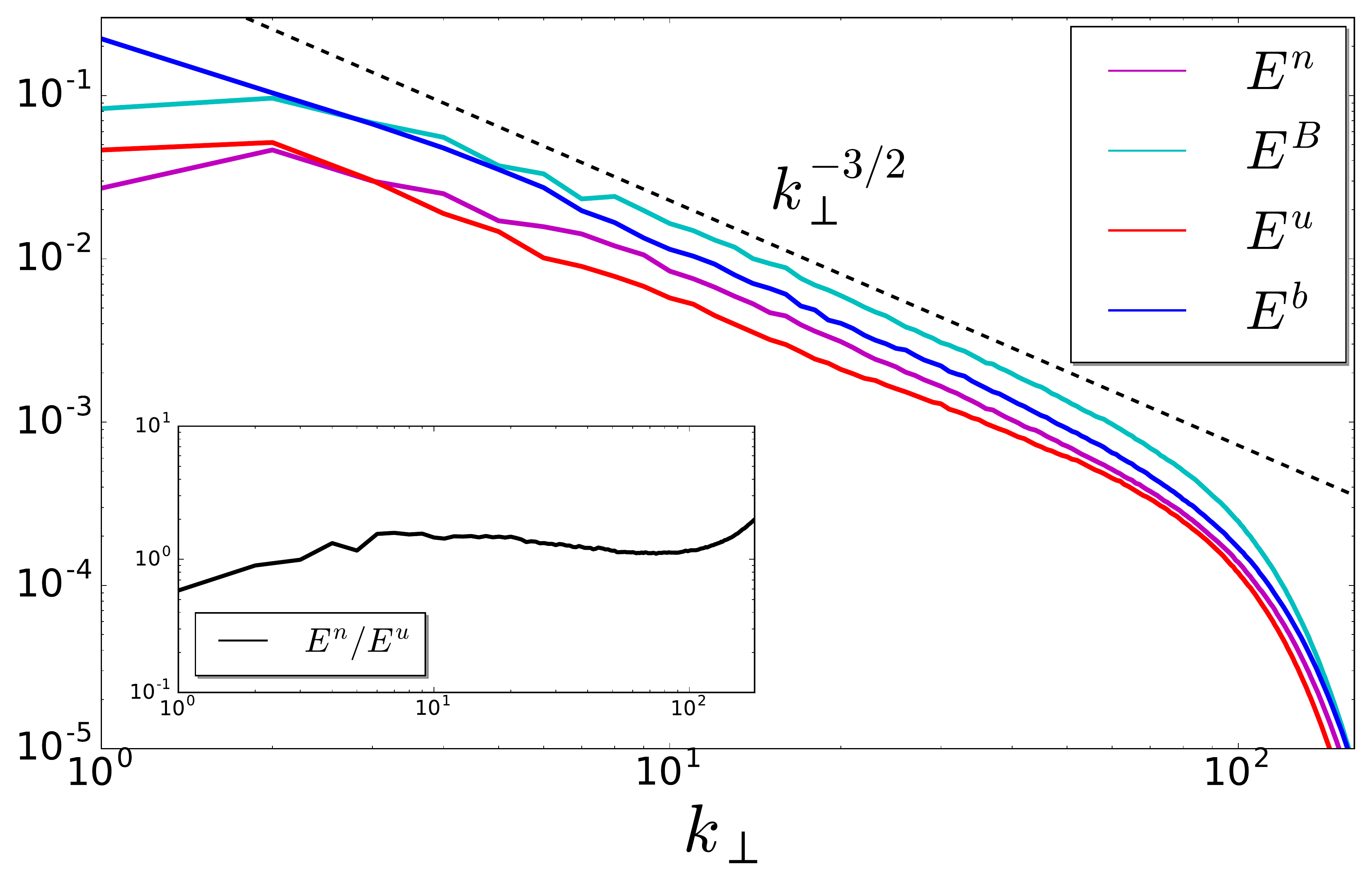}
\end{center}
\vskip-0.5cm
\caption{{\bf Perpendicular cascade.}
Upper panel: snapshot of the density-fluctuation field $\dn$ 
in the plane $(x,y)$ perpendicular to the mean magnetic field.
Lower panel: perpendicular spectra of the Alfv\'enic 
(velocity $E^u$ and magnetic $E^b$), 
density ($E^n$) and magnetic-field-strength ($E^B$) fluctuations; 
inset: ratio of the density to Alfv\'enic-velocity spectra, $E^n/E^u$. 
The Alfv\'enic spectra are normalized to the total mean Alfv\'enic energy,  
compressive spectra to the total mean free energy~$W$ (\eqref{eq:W}).}
\label{fig:perp_cascade}
\vskip-0.25cm
\end{figure}

\subsection*{Perpendicular spectra}
Let us first establish that we are dealing with a system that exhibits some familar features 
of turbulence. 

A snapshot of the density-fluctuation field in the plane perpendicular to the background 
magnetic field is shown in \figref{fig:perp_cascade}, together with time-averaged 
perpendicular wavenumber spectra of the density $\dn$ ($E^n$), magnetic-field strength 
$\dB$ ($E^B$), and of the Alfv\'enic fluctuations of velocity $\vu$ ($E^u$) 
and magnetic field $\vb$ ($E^b$). 
The spectral slope of the latter follows quite well the $\kperp^{-3/2}$ scaling 
expected for RMHD turbulence \cite{boldyrev06,mallet17a,maron01,chenmallet11,perez12}. 
To a good approximation, the 
compressive fluctuations' spectra track the Alfv\'enic-velocity spectrum (see 
inset of \figref{fig:perp_cascade}), as one might expect an undamped passive scalar  
to do \cite{obukhov49,corrsin51}. We do not observe a significant steepening 
of the compressive spectra compared 
to the Alfv\'enic ones, which would have had to happen had there been Landau damping 
of compressive fluctuations depleting their energy cascade at each scale \cite{sch16}. 
This is circumstantially consistent with an inhibition of Landau damping.
It is also essentially consistent with what is observed in the solar wind 
\cite{celnikier83,celnikier87,marsch90,bershadskii04,hnat05,kellogg05,chen11}.

\begin{figure}[t]
\begin{center}
\includegraphics[width=1.0\linewidth]{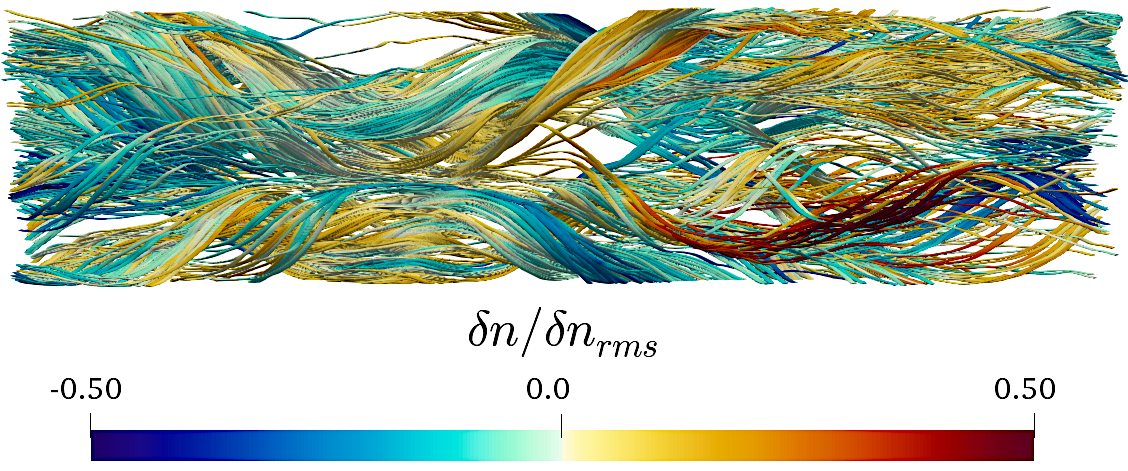}\\
\includegraphics[width=0.9\linewidth]{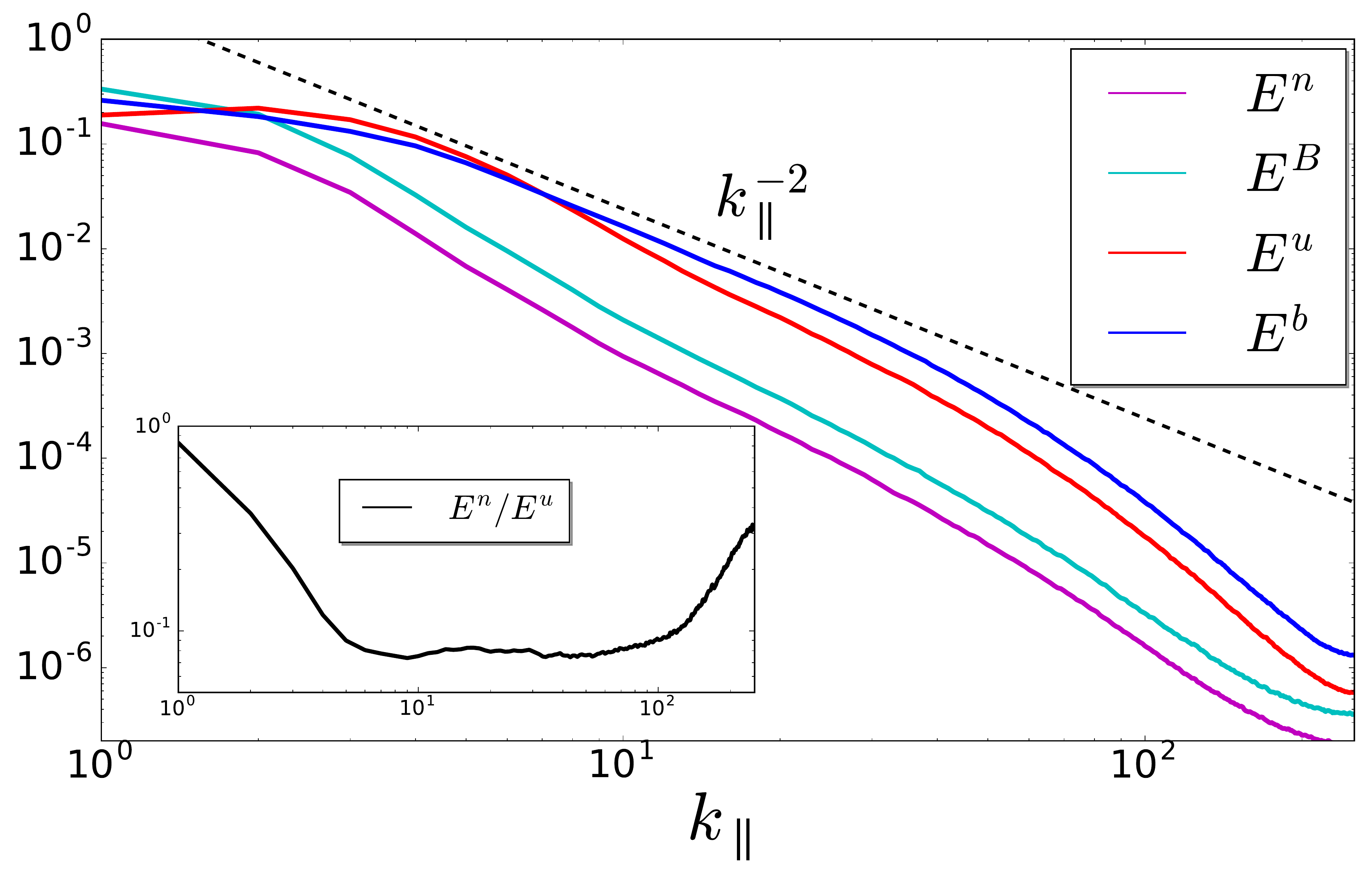}\\
\includegraphics[width=0.85\linewidth]{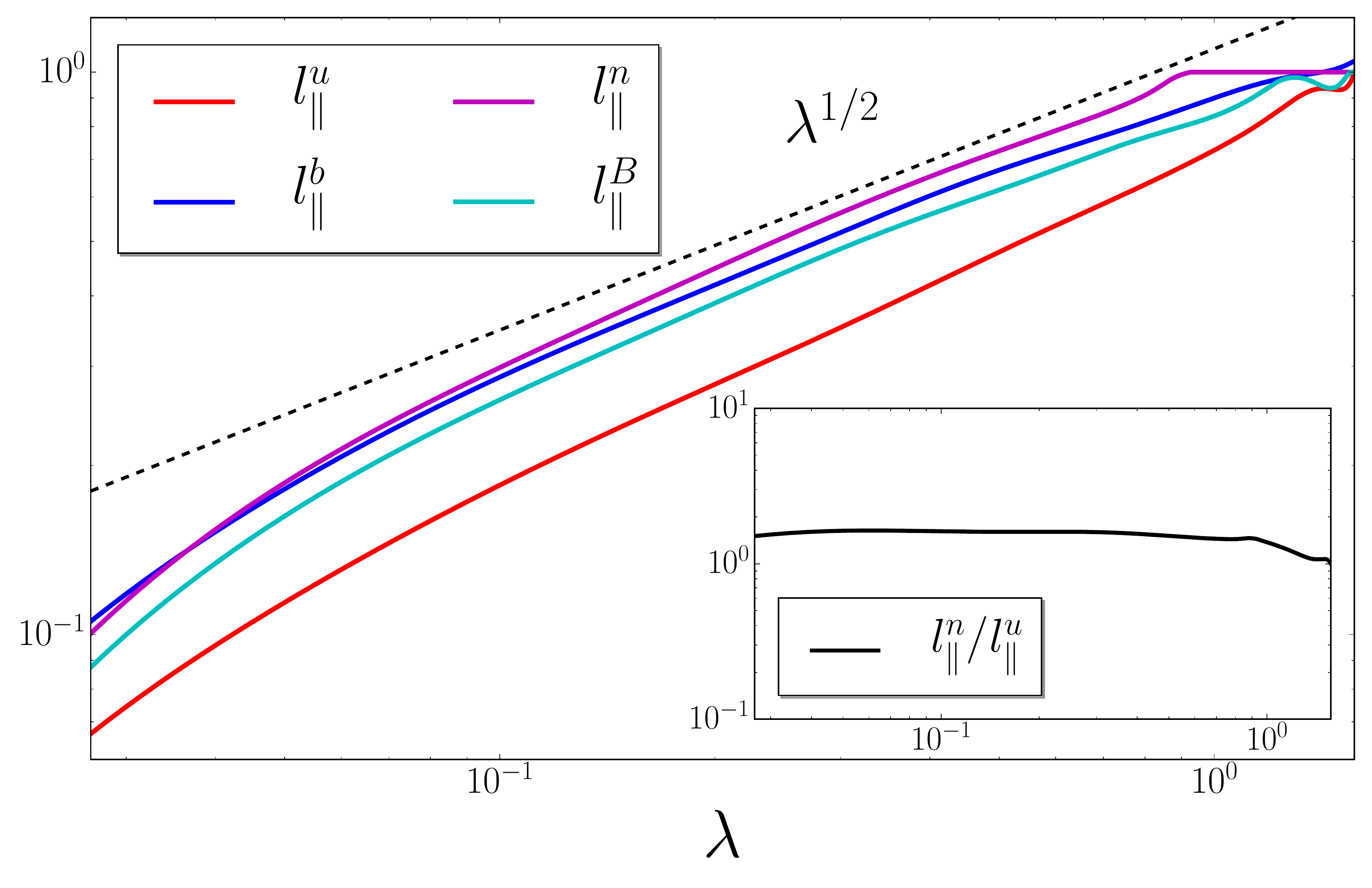}\\
\end{center}
\vskip-0.7cm
\caption{{\bf Parallel cascade.}
Upper panel: snapshot of a typical set of magnetic-field lines, 
with color showing the density-fluctuation field. 
The $z$ dimension of the box is lengthened by the factor of 4 for better viewing 
(formally, in KRMHD, it is arbitrarily longer than the perpendicular size of the box). 
Middle panel: parallel spectra of the Alfv\'enic, density and field-strength fluctuations, 
measured along perturbed field lines; 
inset: ratio of the density to Alfv\'enic-velocity spectra, $E^n/E^u$. 
The normalizations are the same as in \figref{fig:perp_cascade}. 
Lower panel: parallel coherence length $\lpar$ of Alfv\'enic, density 
and field-strength fluctuations as a function 
of their perpendicular coherence length $\lambda$; 
inset: ratio of the parallel coherence length
of the density to that of the Alfv\'enic velocity. 
Here $\lpar(\lambda)$ is calculated via the second-order correlation function, 
by the method described in \cite{mallet15}, 
but following the perturbed field lines with higher-order accuracy, as described in the text.}
\label{fig:par_cascade}
\vskip-0.6cm
\end{figure}

\subsection*{Existence of parallel cascade}
Critically balanced Alfv\'enic fluctuations become progressively 
more anisotropic at finer scales \cite{GS95,GS97}, with parallel and 
perpendicular correlation scales related by 
$\lpar \propto \lperp^{1/2}$ \cite{boldyrev06,mallet15,mallet17a}.  
Consequently, their parallel spectrum is expected to scale as $\kpar^{-2}$ 
\cite{GS97,beresnyak15,mallet17a} (and indeed does in the solar wind 
\cite{horbury08,podesta09aniso,wicks10,chenmallet11,chen16}). 
The question that we address here is whether 
the kinetic field $g$ advected by them ``inherits'' this parallel cascade.

Because $\lpar\gg\lperp$, parallel 
correlations can only be measured correctly  
along the {\it fluctuating} field \cite{cho00,maron01,chenmallet11}. We identify the field 
lines by tracing a set of them from $100,000$ randomly chosen points in the simulation domain 
at a given instant in time, 
record the values of fluctuating quantities as functions of 
the distance along each field line, and then calculate parallel 
spectra (we use a fourth-order Runge-Kutta method to integrate the field lines 
and cubic spline interpolation to determine values of the fluctuating fields on them).

The results from this analysis are shown in 
\figref{fig:par_cascade} alongside an instantaneous snapshot of field lines, 
in which they are ``painted'' with the 
values of the density-fluctuation field. One can see that the 
field lines wander widely across the domain. 
The resulting parallel wavenumber spectra are steeper than the 
perpendicular wavenumber spectra, as expected. The $\kpar^{-2}$ scaling 
is well satisfied for $E^b$, whereas the Alfv\'enic-velocity spectrum $E^u$ is 
a little steeper. The compressive spectra closely track $E^u$ (see inset), 
confirming the existence of a ``parallel cascade'' of these passive fields 
(contrary to what was believed by \cite{sch09} on the basis of linearity 
of \eqref{eq:gb} in a suitably well-behaved Lagrangian frame). 
The lower panel of \figref{fig:par_cascade} reinforces this conclusion:
the parallel coherence lengths of the compressive fluctuations scale as 
$\lambda^{1/2}$, as do those of the Alfv\'enic fluctuations. 

The existence of a parallel cascade is an important finding of the present study.
As we have argued above, it occurs because magnetic fields reconnect at every scale 
over a time comparable to the correlation time associated 
with this scale \cite{eyink15,eyink18,boozer18}
and thus cannot preserve their identity over more than one parallel correlation 
scale of the Alfv\'enic turbulence---thus, any density perturbation that extends along 
a field line will be broken up and decorrelated on the same parallel scale (hence 
the tracking of the Alfv\'enic spectra by the compressive ones). 

It is the presence 
of the parallel cascade that makes the question of the efficiency of Landau damping relevant: 
its rate $\sim|\kpar|\vth$ will be of the same order as the Alfv\'en frequency 
$\kpar\vA$ (taking $\beta_i=1$) and so, in a critically balanced turbulence, of the 
same order as the nonlinear cascade rate at every scale. It therefore {\em a priori} 
matters as an energy-removal channel and the discovery that it in fact does not is 
a nontrivial one. 

\subsection*{Phase-space spectra and dissipation}
Let us now complete the characterisation of the free-energy distribution in 
phase space by studying the structure of the fluctuations of $g$ in velocity space, 
in terms of its Hermite spectra. 
Besides helping us build our case for fluidization of kinetic turbulence, 
these are of interest in the context of the rapid recent advances in both 
instrumentation and computing, meaning that they can now be measured 
directly both in space \cite{servidio17} and in fully kinetic simulations 
\cite{cerri18,pezzi18,kawazura18ie}.  

\begin{figure}[t]
\includegraphics[width=1.0\linewidth]{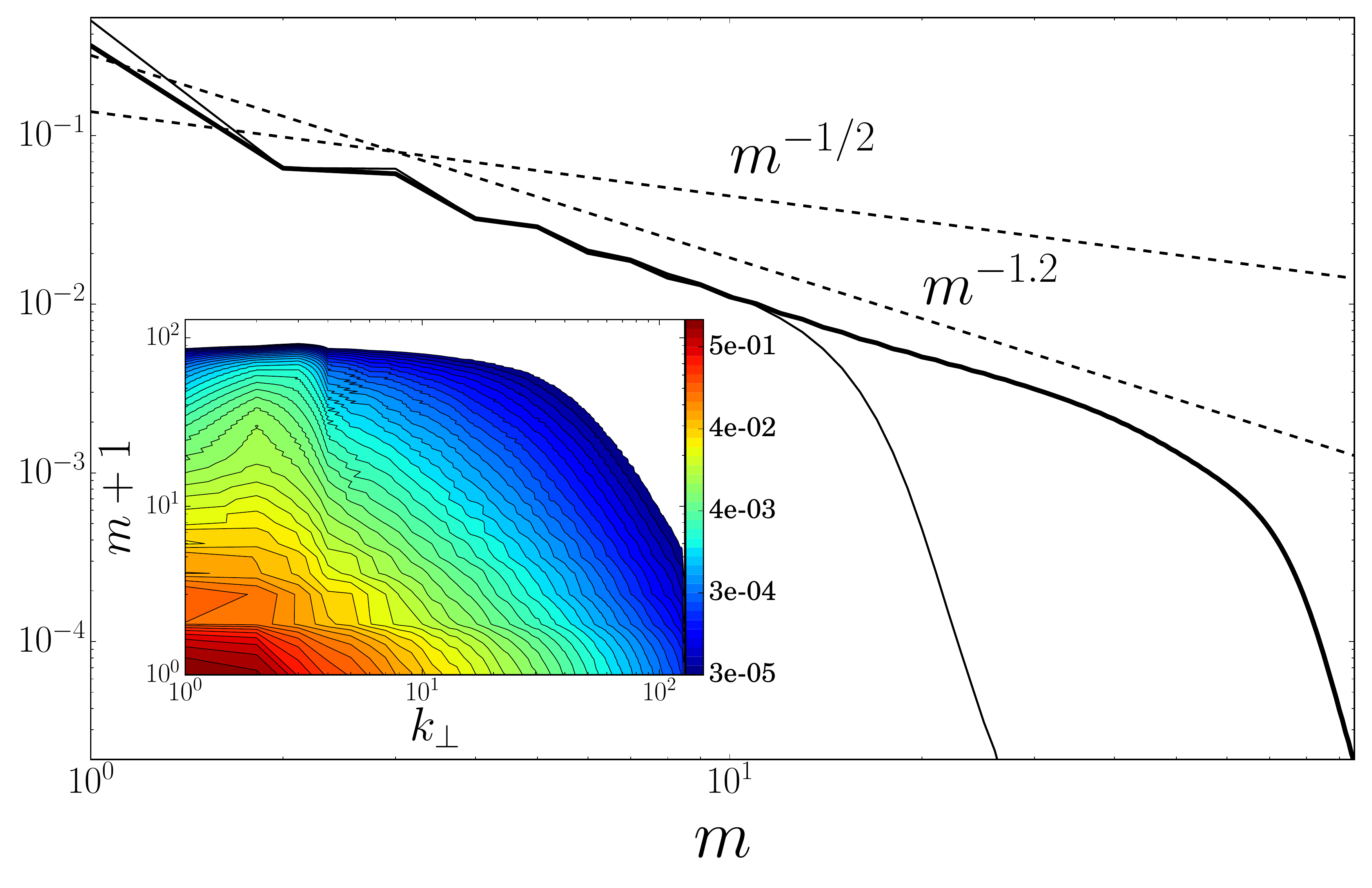}
\caption{{\bf Phase-space cascade.}
Hermite spectrum (integrated over wavenumbers) of the kinetic field $g^{(1)}$ advected by Alfv\'enic 
turbulence according to \eqref{eq:gb}: the thick line shows the spectrum for a run with spatial 
resolution $256^3$ and $M=128$ Hermite moments; the thin line is for a run with spatial 
resolution $512^3$ and $M=32$ Hermite moments. The slope $m^{-1/2}$ associated with 
linear Landau damping \cite{zocco11,kanekar15} is shown for reference.
The normalization of the spectrum is the same as in \figref{fig:perp_cascade}.
Inset: spectrum of $g^{(1)}$ in the phase space $(\kperp,m)$ (integrated over $k_z$).}
\label{fig:g_spec}
\end{figure}

\begin{figure}[t]
\includegraphics[width=1.0\linewidth]{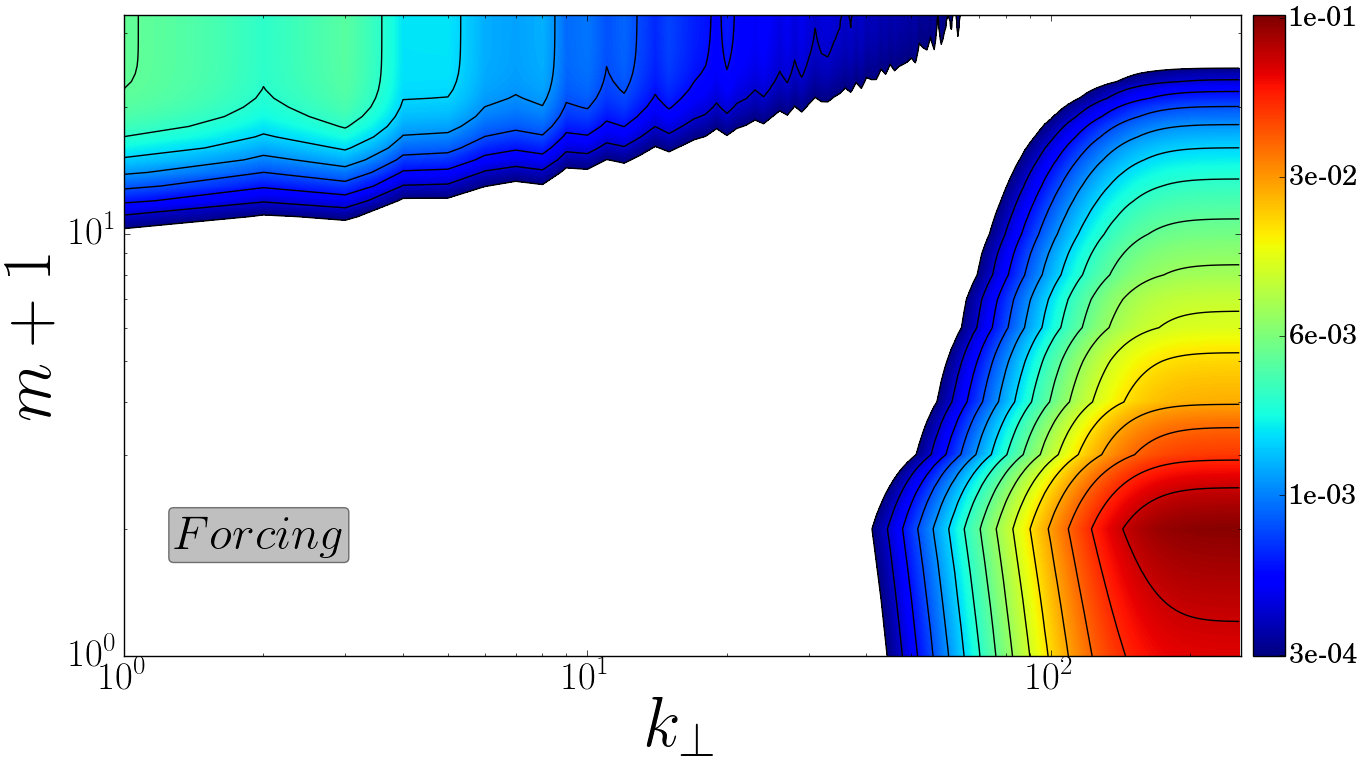}
\caption{Dissipation rate of the free energy $W$ (defined by \eqref{eq:W}) 
of the kinetic field $g^{(1)}$, normalized to its total injection rate, 
as a function of the Hermite number $m$ and perpendicular wavenumber $\kperp$.  
Dissipation is important 
only for large $m$ (hypercollisional) or large $\kperp$ (hyperviscous). 
In the inertial range, the free energy is well conserved. 
The range of $(\kperp,m)$ into which energy is injected by forcing is shown as grey box. 
Some of the energy injected at $m=1$ and 
low $\kperp$ is dissipated by collisions, but the dissipation is most intense at 
high $\kperp$ and low $m$.} 
\label{fig:diss_spec}
\end{figure}

In the absence of background Alfv\'enic turbulence, 
the time- and space-averaged Hermite spectrum of a forced and Landau-damped 
kinetic field is $\la|g_m|^2\ra \propto m^{-1/2}$ \cite{zocco11,kanekar15}. 
This corresponds to a constant finite free-energy flux from small to large Hermite numbers. 
As a consequence, there is a dissipative anomaly associated with the collision operator: 
in the limit of vanishing collisionality, the collisional dissipation stays finite, enabling 
the removal of free energy from the system at a collisionality-independent 
phase-mixing rate. With turbulence, we find the spectrum to be a little steeper 
than $m^{-1}$, as shown in \figref{fig:g_spec}. Collisional dissipation rate 
associated with such a spectrum would vanish in the limit of small 
collisionality \cite{adkins18}, 
indicating that refinement of scales in velocity space is not sufficient to 
process into heat the finite free energy injected by forcing, and thus that the principal 
cascade route must be via smaller {\it spatial} scales. 
Indeed, \figref{fig:diss_spec}, where we show the combined collisional and hyperviscous dissipation 
in the $(\kperp,m)$ phase space, confirms that the majority of the free energy 
is thermalized at small spatial scales, not at high~$m$. 

\subsection*{Free-energy fluxes}
Finally, having collected all this circumstantial (but observationally testable)
evidence for the ``fluidization'' 
of kinetic turbulence, let us return to \figref{fig:flux}, where the fluxes 
of free energy in the $(\kperp,m)$ space are plotted.  
Comparison of the fluxes towards smaller scales in velocity space (higher Hermite numbers) 
with the fluxes towards smaller scales in position space (higher wavenumbers) provides 
the most direct numerical evidence supporting the claim that, in the inertial range, 
Landau damping is suppressed rather than enhanced. 

The compressive free energy, $W$, defined by \eqref{eq:W}, 
is conserved in the inertial range of our simulations. 
Indeed, forcing is implemented spectrally (in $m$ and $\kperp$) and is, therefore, 
exactly zero outside of a few lowest modes. 
The dissipation rate (shown in \figref{fig:diss_spec}) is negiligibly small away from the highest 
resolved $m$ and $\kperp$. For most $m$ and $\kperp$ modes, there are, therefore, neither sinks 
nor sources of free energy other than coupling to other modes. 
Fluxes from mode to mode can then be defined in such a way that 
the free energy in a wavenumber shell $|\vkperp'|=\kperp$ and at Hermite number $m$, 
$W_m(\kperp)=\sum_{k'_z}\sum_{|\vkperp'|=\kperp}(1+\alpha\,\delta_{m,0})|g_{m,\vk'}|^2/2$, satisfies, 
away from the forcing and dissipation,  
\beq
\frac{\dd W_m(\kperp)}{\dd t} = - \lt[\Gamma_m(\kperp) - \Gamma_{m-1}(\kperp)\rt]
-\frac{\dd\Pi_m(\kperp)}{\dd\kperp}, 
\eeq
where $\Gamma_m(\kperp)$ is the Hermite flux and $\Pi_m(\kperp)$ is the Fourier flux. 
The Hermite coupling is essentially local 
(it involves only neighboring Hermite numbers $m-1$, $m$ and $m+1$), 
and so it is easy to read off Hermite fluxes from \eqref{eq:gm}: 
\begin{align}
\nonumber
&\Gamma_m(\kperp) = \vth\sum_{m'=0}^m\int\frac{\rmd^3\vr}{V} (1+\alpha\,\delta_{m',0}) [g_{m'}]^=_{\kperp}\\
&\qquad\dpar\lt[\sqrt{\frac{m'+1}{2}}\,g_{m'+1} 
+ (1+\alpha\,\delta_{m',1})\sqrt{\frac{m'}{2}}\,g_{m'-1}\rt],
\label{eq:Gamma}
\end{align}
where the Fourier ring filter applied to a function is defined by
\beq
[g_m]^=_{\kperp}(\vr) = \sum_{k'_z}\sum_{|\vkperp'|=\kperp} e^{i\vk'\cdot\vr} g_{m,\vk'}. 
\eeq
Fluxes in Fourier space can be nonlocal, but are commonly defined and studied 
nonetheless \cite{verma04}:
\beq
\label{eq:Pi}
\Pi_m(\kperp) = (1+\alpha\,\delta_{m,0})
\frac{2\pi}{\sqrt{L_x L_y}}
\int\frac{\rmd^3\vr}{V}\,[g_m]^>_{\kperp}\vu\cdot\vdperp[g_m]^{\le}_{\kperp},
\eeq
where $L_xL_y=(2\pi)^2$ is the cross-section of the box and 
the high- and low-pass Fourier filters are defined by 
\begin{align}
&[g_m]^>_{\kperp}(\vr) = \sum_{k'_z}\sum_{|\vkperp'|>\kperp} e^{i\vk'\cdot\vr} g_{m,\vk'},\\
&[g_m]^\le_{\kperp}(\vr) = \sum_{k'_z}\sum_{|\vkperp'|\le\kperp} e^{i\vk'\cdot\vr} g_{m,\vk'}.
\end{align}
While the two-dimensional 
phase-space flux $(\Pi,\Gamma)$ is not a unique quantity, defined up to 
arbitrary circulations in the $(\kperp,m)$ space, it is a useful one and, absent 
any obviously spurious such circulations, it gives a good representation of the paths that 
free energy takes to cross the gap between the forcing and dissipation scales. 

It is $(\Pi,\Gamma)$ vs.\ $(\kperp,m)$, as defined by \eqsand{eq:Pi}{eq:Gamma}, 
that is plotted in \figref{fig:flux}. 
On average, at forcing scales (low $\kperp$), a significant amount of free energy flows 
directly to high $m$. This flux is restricted to the forcing band of $\kperp$ modes, 
because in this range, the nonlinear interactions have not yet managed to set 
up a return echo flux. This is in sharp contrast with the situation in the inertial range, 
where the energy flows mostly to small spatial scales. Recall that in the inertial range, 
the parallel cascade of compressive fluctuations should make Landau damping {\it stronger} 
in the presence of turbulence than otherwise, because free-streaming particles have less 
distance to travel to smooth out spatial fluctuations. Recall also that this parallel cascade 
is critically balanced, which means that Landau damping should be 
able to ``keep up'' with nonlinear, spatial mixing at every scale in the inertial range. 
In reality (as represented by the simulations), Landau damping 
is manifestly rapid enough to be important at forcing scales but (on average) 
far weaker than perpendicular spatial mixing in the inertial range, so it is reasonable 
to conclude that there is another process that is cancelling on average 
the flux to high~$m$. 
This is the most explicit signature of stochastic echoes that can be diagnosed in these 
simulations.\footnote{For electrostatic kinetic turbulence, where $\vb=0$, it is 
possible to decompose $g_m$ explicitly into phase-mixing and anti-phase-mixing 
components and thus calculate directly the ``forward'' and ``backward'' free-energy 
fluxes in Hermite space \cite{sch16,adkins18}. 
However, this decomposition involves the sign of $\kpar$, 
which in our case would have to be calculated with respect to the perturbed, turbulent 
magnetic field. We do not know how to make such a calculation mathematically 
rigorous and thus do not use this decomposition.}
%It is useful at this point to glance back at \figref{fig:diss_spec} and note that 
%free energy injected in $m=1$ is largely dissipated still in $m=1$, with relatively 
%little leakage to other $m$'s. 

The conclusion from \figref{fig:flux} is that, in the inertial range, individual Hermite 
moments are, in effect, energetically insulated from each other, on average. 
The compressive turbulence is ``fluidized''. 

\section*{Discussion}

\subsection*{Summary}
We have presented a case for effective fluidization of compressive fluctuations in 
collisionless (or weakly collisional) plasma turbulence. Compressive fluctuations 
are advected passively by the ambient Alfv\'enic turbulence and turn out to inherit 
its parallel structure (\figref{fig:par_cascade}). This means that their phase-mixing 
(Landau-damping) rate $\sim|\kpar|\vth$ is in principle comparable to the 
frequency $\kpar\vA$ of the Alfv\'enic motions and, therefore, in a critically 
balanced turbulence \cite{GS95,GS97}, to the rate at which these motions push the compressive 
free energy to smaller perpendicular scales. However, what might have been thought 
an effective dissipation at every scale fails to remove energy efficiently from 
the low-order moments of the ion distribution function (\figref{fig:flux}) and thus steepen the 
inertial-range spectra of, e.g., density and magnetic-field-strength fluctuations. 
Instead, they follow faithfully the spectrum of the advecting Alfv\'enic field 
(\figref{fig:perp_cascade}) as a well-behaved ``fluid'' passive scalar would 
do \cite{obukhov49,corrsin51}. The reason for this turns out to be that 
nonlinear advection effectively nullifies phase mixing (except at the forcing scales), 
with free-energy fluxes in the inertial range largely confined within 
each moment of the distribution function, taking its energy from large 
to small {\em spatial} scales, where it dissipates (\figref{fig:diss_spec}).  
We interpret this behavior 
as resulting from statistical cancellation of phase mixing by plasma echoes. These 
are excited because the nonlinear advection causes phase-mixing modes propagating 
towards smaller velocity-space scales to couple to anti-phase-mixing ones, 
propagating back to larger scales--- the result is a return flux from phase space \cite{sch16}. 

\begin{figure*}[t]
\includegraphics[width=1.0\textwidth]{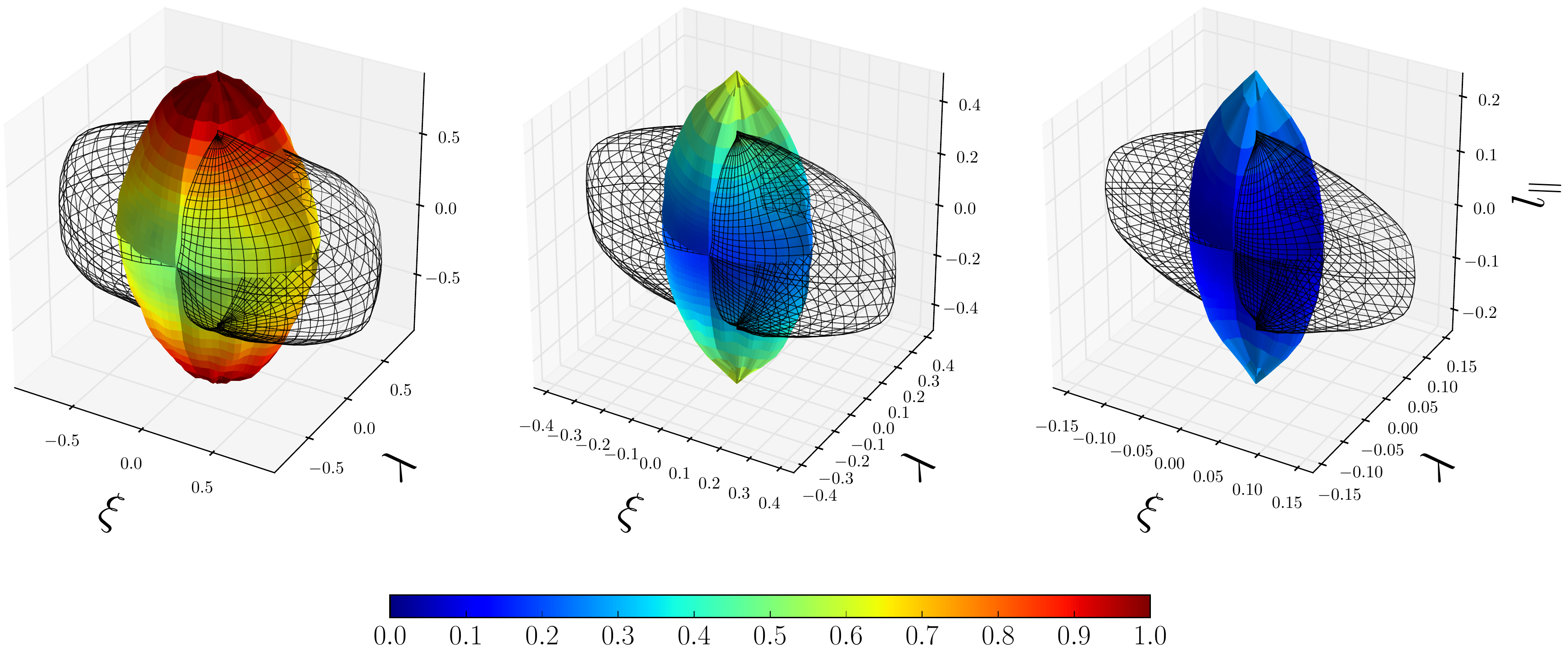}
%\vskip-1cm
\caption{Level sets for the correlation functions of 
the Alfv\'enic magnetic field $\vb$ (black netting)
and perturbed field strength $\dB/B_0$ (color) corresponding to the perpendicular scales 
(from left to right) $\lambda=0.5,0.16,0.05$.
The two other coordinates are the distance $\lpar$ along local mean field 
and the dustance $\xi$ along the perturbed field $\vb$. 
Colors show distance from the center. Note that the parallel and perpendicular 
scales are measured in units of parallel and perpendicular box size, respectively  
($2\pi$ in each direction), which are also the forcing scales, 
but, in KRMHD, the box can be arbitrarily stretched in the parallel direction---so these 
level sets are in fact highly elongated with $\lpar\gg\xi,\lambda$. 
These correlation functions have been constructed following the method 
desribed in \cite{mallet16} and are to be compared, modulo the arbitrary 
elongation factor, with solar-wind measurements in \cite{chen12}.}
\label{fig:3Deddies}
\end{figure*}

\subsection*{Relation to observations in space}
There is a long history of density and field-strength spectra being measured in 
the solar wind and found to have the same slopes as the ambient Alfv\'enic 
turbulence \cite{celnikier83,celnikier87,marsch90,bershadskii04,hnat05,kellogg05,chen11}.
Why they do so in a collisionless plasma has remained a puzzle, if perhaps not 
always a fully appreciated one, in view of the community's instinctive 
preference for fluid models. As solar wind has come to be regarded as a unique 
plasma physics laboratory, compressive fluctuations, under further, more plasma-physics-aware, 
scrutiny, have stubbornly continued to manifest ``fluid'' behavior \cite{verscharen17}.
Ours is the first dedicated attempt to simulate these fluctuations 
drift-kinetically, using equations that are physically appropriate 
for them \cite{sch09}. Our results are manifestly in agreement with what observations 
have shown for many years. 

In the solar wind, 
it has, relatively recently, become possible to probe beyond spectra and diagnose 
three-dimensional structure of local correlations \cite{chen12,verdini18}. The physically 
motivated local basis is the direction of the ``local mean'' magnetic field 
(the correlation length along it is denoted $\lpar$), 
the direction of the field's perturbation $\vb$ at the scale of interest 
(the corresponding correlation length is $\xi$) and the third direction, transverse to 
both (correlation length $\lambda$). The idea is that, if turbulence has 
a tendency towards local alignment between the two Alfv\'enic fields 
\cite{boldyrev06,chandran15,mallet16,mallet17a}, it must turn out that 
statistically not only $\lambda,\xi\ll\lpar$ (anisotropy) but also 
$\lambda\ll\xi$ (alignment). In \figref{fig:3Deddies}, we show 
representative contours of 3D correlation functions (``statistical eddy shapes'') 
calculated in this frame both for Alfv\'enic and compressive fluctuations. 
They are very similar 
to what is measured in the solar wind \cite{chen12}: Alfv\'enic ``statistical eddies'' 
are ``pancakes'', or ``ribbons'', with $\lambda\ll\xi\ll\lpar$, while the compressive 
structures have a more tubular aspect.\footnote{Quantitatively, for the Alfv\'enic fields, we see 
a scaling of the aspect ratio (very approximately) consistent with $\lambda/\xi\propto\lambda^{1/4}$ 
\cite{boldyrev06,mallet16,mallet17a}, whereas for the compressive fluctuations,  
$\lambda/\xi$ has a noticeably shallower scaling with $\lambda$ (we do not 
measure these alignment scalings accurately at the resolution of our simulations). 
Thus, compressive fluctuations do appear to have a local 3D anisotropy and a tendency 
to form ribbons, but a weaker one than Alfv\'enic turbulence exhibits.}  
Thus, it appears that simulated collisionless compressive fluctuations bear significant 
resemblance to the measured ones, in a more detailed way than just having the same 
spectra---a reassuring fact.
   
With the extraordinary velocity-space resolution afforded by the new MMS satellite, 
phase-space spectra have become measurable in space plasmas \cite{servidio17}. 
Thus, the prediction of steep Hermite spectra (\figref{fig:g_spec}) is a falsifiable one. 
More generally, considerations of phase-space turbulence can now be viewed as 
more than just theorizing about ``under-the-hood'' physics, for they
deal with phenomena that are directly observable with available instruments.  

\subsection*{Implications for modelling techniques}
As we mentioned in the introduction, there is a thriving industry of effective fluid 
models of collisionless plasmas, of which the most sophisticated strand is the ``Landau-fluid'' 
closures \cite{hammett90,hammett92,dorland93,beer96,snyder97,snyder01,passot04,sulem15,tassi16}. 
Their underlying idea is to assume that Landau damping removes free energy from low 
to high Hermite moments as effectively 
in a nonlinear system as in a linear one. This might appear 
to be the exact opposite of the main conclusion of this work. However, Landau-fluid models 
have consistently been found to work better when more---but 
not necessarily many more---moments are kept compared to the standard fluid approximation. 
It might be argued that the art of crafting a good Landau-fluid model is precisely to do it 
in such a manner as to capture the effect of the echoes within a minimal set of Hermite moments 
while setting the boundary (closure) condition at the maximum retained $m$ so as not to introduce 
or divert free-energy flows in a spurious way. With the echo effect and fluidization 
now explicitly part of one's intellectual vocabulary, one might hope to revisit this task 
with renewed vigor, purpose and insight. 

\subsection*{An implication for astrophysical theory}
One cannot either adequately enumerate or, indeed, anticipate all of the 
instances across the vast canvas of plasma astrophysics where the nature 
of collisionless plasma turbulence may prove to be of interest. We wish 
to highlight one problem that has a long history \cite{rees82,narayan95,quataert99} 
and has recently seen 
a burst of activity (see \cite{kawazura18ie} and references therein). 
It is a long-standing question in the theory of matter accretion onto black holes 
whether, and to what degree, plasma turbulence that is excited in the accretion disk 
by instabilities driven by the Keplerian shear and helps enable accretion 
by transporting angular momentum, can be thermalized preferentially on ions, rather 
than electrons, or vice versa. This has implications for the relative amounts of 
energy radiated out by electrons (and thus observed) vs.\ swallowed by the black hole 
as mass (ions) is sucked in, as well as for observational signatures of disks 
and their jets \cite{ressler17,chael18}. Turbulent energy is split into Alfv\'enic and compressive 
cascades at MHD scales \cite{lithwick01,sch09}. Whereas at asymptotically low $\beta_i$, 
one can show that all of the compressive free energy 
must always thermalize into ions (this follows from the equations derived in \cite{zocco11}), 
how it is partitioned 
between ions and electrons at finite and high $\beta_i$ 
is a nontrivial question decided by the dynamics at the ion Larmor 
scale \cite{howes11prl,told15,kawazura18ie,parashar18}---and so how much of it 
arrives to this scale without being dissipated %(necessarily, onto ions) 
on its way through the inertial range is important. 
Our findings indicate that, at least at $\beta_i\sim1$, most of it does. 
Since the theory (or even a reliable modelling prescription) 
of energy partition in plasma turbulence is still to be developed, this is a useful 
factual constraint to have. 

\subsection*{Implications for general (plasma) physics}
The peregrinations and rearrangements 
of energy through a system's phase space is a recurrent motif of theoretical physics. Turbulence 
theory is explicitly constructed to describe the energy's thermalization routes, which 
bridge the usually vast separations between its injection and dissipation 
scales, producing rich, multiscale nonlinear structure in the process. In weakly collisional 
plasmas, these energy-transfer 
routes lie in a 6D phase space, with velocity-space refinement (phase mixing) 
of the particles' distribution functions in general as effective as spatial mixing 
in accessing dissipation mechanisms \cite{sch09,eyink18}. It is perhaps noteworthy 
that, in the case of inertial-range turbulence of a magnetized plasma, one of these 
forms of mixing turns out unambiguously to be the winner: spatial advection outperforms 
phase mixing and makes a collisionless plasma resemble a collisional fluid. Those 
who believe in the universality of nonlinear dynamics might be pleased by such an outcome. 
For plasma physicists, this is a sobering reminder that collisionless dissipation 
processes that make our subject so intellectually distinctive are not irreversible until they 
are consummated by collisional entropy production---and an intriguing demonstration 
that nonlinear effects can sometimes hinder them in favor of more ``fluid-like'' 
entropy-production mechanisms.

\acknow{We are grateful to T.~Adkins, T.~Antonsen, A.~Beresnyak, 
F.~Califano, B.~Chandran, S.~Cowley, P.~Dellar, G.~Hammett, E.~Highcock, Y.~Kawazura, N.~Loureiro, A.~Mallet, J.~Parker, C.~Staines, and L.~Stipani  
for many useful interactions on topics related to this project. 
This work was made possible thanks to the HPC resources of TGCC made available by GENCI (project A0010510117). 
W.D.\ and A.K.\ received support from US DoE (grants DE-FC02-04ER54784 and DE-FG02-93ER54197).
The work of A.A.S.\ was supported in part by grants from 
UK STFC (ST/N000919/1) and EPSRC (EP/M022331/1).
All authors gratefully acknowledge the hospitality of the Wolfgang Pauli Institute, Vienna, where 
some of this work was done.}

\showacknow % Display the acknowledgments section

% \pnasbreak splits and balances the columns before the references.
% If you see unexpected formatting errors, try commenting out this line
% as it can run into problems with floats and footnotes on the final page.
%\pnasbreak

\bibliography{bib_JPP}{}

\end{document}